\documentclass[submission,copyright,creativecommons]{eptcs}
\providecommand{\event}{FLACOS 2011} 
\usepackage{breakurl}             
\usepackage{color,graphicx,algorithmic,amsfonts,amssymb,fancyhdr,bigdelim,multirow,bigstrut,wrapfig,float,fancyvrb,xspace}
\usepackage{display-figure}
\usepackage{amsthm}
\usepackage{amsfonts}
\usepackage{amsmath}
\usepackage{amssymb}
\usepackage{graphicx,url}
\usepackage{amsfonts}
\usepackage{amsmath}
\usepackage{amssymb}
\usepackage{graphicx,url}
\usepackage{listings}
\usepackage{ dsfont,latexsym,longtable,epsfig,color}
\usepackage[latin1]{inputenc}
\usepackage{fontenc}
\usepackage{float}
\usepackage{stmaryrd}
\usepackage{mathpartir}
\usepackage{txfonts}
\usepackage{display-figure}
\usepackage{verbatim}
\usepackage{amssymb}
\usepackage{subfig}
\usepackage{mathtools}
\usepackage{andrew-macros}
\usepackage{algorithmic,fancyhdr,bigdelim,multirow,bigstrut,wrapfig,fancyvrb,xspace}

\newtheorem{example}{\textsl{Example}}[section]

\title{Distributed System Contract Monitoring}
\author{Adrian Francalanza \qquad Andrew Gauci \qquad Gordon J. Pace
\institute{Department of Computer Science, University of Malta
\thanks{The research work disclosed in this publication is partially funded by the Strategic Educational Pathways Scholarship Scheme (Malta). The scholarship is part financed by the European Union European Social Fund.}\\
}
\email{\{adrian.francalanza|agau0006|gordon.pace\}@um.edu.mt}
}
\def\titlerunning{Distributed System Contract Monitoring}
\def\authorrunning{A. Francalanza, A. Gauci \& G.J. Pace}
\begin{document}
\maketitle


\begin{abstract}

The use of behavioural contracts, to specify, regulate and verify systems, is particularly relevant to runtime monitoring of distributed systems. System distribution poses major challenges to contract monitoring, from monitoring-induced information leaks to computation load balancing, communication overheads and fault-tolerance. We present \MDPi, a location-aware process calculus, for reasoning about monitoring of distributed systems.  We define a family of Labelled Transition Systems for this calculus, which allow formal reasoning  about different monitoring strategies at different levels of abstractions. We also illustrate the expressivity of the calculus by showing how contracts in a simple contract language can be synthesised into different \MDPi\ monitors.
\end{abstract}

\section{Introduction}
\label{sec:introduction}

As systems continue to grow in size and complexity, the use of behavioural contracts  is becoming crucial in specifying, regulating and verifying correctness.
Various notions of contracts have been used, but most prevailing variants enable the regulation of the behaviour of a system, possibly with consequences in case of violations. Such contracts can then be used in multiple ways, from system validation and verification, to 
conflict analysis of the contract itself. One important use of contracts is in runtime monitoring:  system traces are analysed at runtime to ensure that any contract violating behaviour is truncated before it leads to any further consequences, possibly applying reparations to recover from anomalous states.

More and more systems are deployed in a distributed fashion, whether out of our  choice or necessity.  
Distribution poses major design challenges for runtime monitoring of contracts, since monitors themselves can be distributed, and trace analysis can be carried out remotely across location.  This impacts directly various aspects of the system being monitored, from the security of sensitive information, to resource management and load balancing, to aspects relating to fault tolerance.  Various alternative solutions have been presented in the literature, from fully orchestrated solutions where monitors are located at a central location, to statically distributed monitors where the contract monitor is statically decomposed into different components hosted at the location where system traces are generated.


The primary contribution of this paper is a unified formal framework for studying different monitoring strategies. We present a location-aware calculus supporting explicit monitoring as a first class entity, and internalising behavioural traces at the operational level rather than at a meta-level. We show the expressivity of the calculus by using it to model different distributed system monitoring strategies from the literature.  We also present a novel architecture in which contract monitors \emph{migrate} across locations to keep information monitoring local, while limiting remote monitor instrumentation in certain situations.
The versatility of the contract-supporting calculus is later illustrated by showing how it can model different instrumentation strategies. In particular, we show how behavioral contracts expressed using regular expressions can be automatically translated into monitors of different monitoring strategies.

The paper is organised as follows. In section \ref{sec:distmon}, we outline the contract monitoring strategies for distributed systems from the literature.  We present a monitoring distributed calculus in sections \ref{sec:language} and \ref{sec:monitoring-semantics}, and illustrate its use for monitoring behavioural contracts expressed as regular expressions in section \ref{sec:regexp}. We discuss related work and conclude in section \ref{sec:conclusions}.

\section{Monitoring Distributed Systems}
\label{sec:distmon}

Monitoring distributed systems is distinct from monolithic monitoring.  These systems are usually characterised by the absence of a global clock when ordering events across location boundaries.  They often consist of autonomous, concurrently executing subsystems communicating through message passing, each with its local memory, where communication across subsystems is considerably slower than local communication. The topology of such systems may sometimes change
at runtime through the addition of new subsystems or the communication of private channels.   Most internet-based and service-oriented systems, peer-to-peer systems and Enterprise Service Bus architectures \cite{ESB} are instances of such systems.

These characteristics impinge on contract monitoring. For instance, the absence of a global clock prohibits precise monitoring for  consequentiality contracts across locations \cite{Lamport:1977:PCM:1313313.1313439}. Distribution impacts on information locality; subsystem events may contain confidential information which must not be exposed globally  thereby  requiring local monitoring.  The possibility of distributing monitors also introduces concerns for monitor load balancing and monitor replication for fault tolerance.

Whether monitored contracts are known at compile time or else become known at runtime also affects distributed monitoring. Static contracts, ones which are fully known at compile time, are typically not expressive enough for distributed systems with dynamic topologies. Dynamic contracts, ones which are 
partially known at compile time, tend to be more appropriate for such systems. They are found 
in
 intrusion detection 
 \cite{Denning87anintrusion-detection}, where suspicious user behaviour is learnt at runtime, and in systems involving service discovery, where the chosen service may come with a fixed or negotiated contract upon discovery.


\subsection{Classifying Distributed System Monitoring Approaches}

Existing approaches for distributed system monitoring can be broadly classified into two categories: \emph{orchestration}-based  or \emph{choreography}-based.  Orchestration-based approaches relegate monitoring responsibility to a central monitor overhearing all necessary information whereas choreography-based typically distribute monitoring across the subsystems.  Orchestration, used traditionally in monolithic systems, is relatively the simplest strategy and its centralisation facilitates the handling of dynamic contracts.  The approach is however susceptible to data exposure when contacts concern private information; it also leads to considerable communication overhead across locations, and poses a security risk by exposing the monitor as a central point of attack. By contrast, choreography-based approaches push verification locally, potentially minimising data exposure and communication overhead. Communication between localised monitors is typically substantially less than that induced by the remote monitoring of a central monitor.  Choreography is however more complex to instrument, as contracts need to be decomposed into coordinating local monitors, is more intrusive, burdening monitored subsystems with additional local computation, and is applicable only when the subsystems allow local instrumentation of monitoring code.  Choreographed monitors are also instrumented upfront, which may lead to redundant local instrumentation in the case of consequential contracts; if monitoring at location $k$ is dependent on verification at location $l$, and the check at $l$ is never satisfied, upfront monitor instrumentation at $k$ is never needed.

Static orchestration 
verifying pre-determined contracts is a common approach, \eg \cite{RTML}, where web-service compositions are monitored in an orchestrated fashion.  By contrast, \cite{WISEMon:WICT2010} uses orchestration to monitor for dynamic properties: web services are centrally monitored against BPMN workflow specifications, facilitating the verification of contracts (representing system properties) discovered 
at runtime \textit{on-the-fly}.  Extensive work has also been done in static choreography monitoring \cite{EfficientDec,drusinksy,SOS-ESB,GEM,conf/rv/ZhouSLL09}, where communication overhead is mitigated by breaking up contracts into parts which can be monitored independently and locally, synchronising between the monitors only when necessary.  However, to the best of our knowledge, these approaches cannot fully handle dynamic contracts with runtime contract decomposition and distribution, nor do they tackle monitor-induced data exposure.

An alternative approach is that of using \emph{migrating monitors}, which adequately supports dynamic contracts whilst still avoiding orchestration; in particular, it limits instrumentation of distributed monitors in cases were monitoring is dependent on computation. Using this approach, monitors reside where the immediate confidential traces reside, and migrate to other subsystems, possibly discovered at runtime, when information from elsewhere is required \ie on a \emph{by-need} basis.  This enhanced expressivity also permits support for dynamic topologies and contracts learnt at runtime.

\begin{example} \label{eg:general-example}
Consider the hospital system contract:
\begin{quote}
  A nurse will have access to a patient's records after requesting
  them, as long as his or her request is approved by a doctor assigned
  to the patient.
\end{quote}
We assume that (i) the nurse requests (and eventually accesses) the patient's data from a handheld device, (ii) the information about which doctors have been assigned to which patients resides at the central site, and (iii) the patient's information is stored on the doctors' private clinic systems, where doctors can also allow nurses permission to access patients' data.

A migrating monitor starts on the nurse's system; upon receiving a  patient-information request, it migrates to the hospital system, decomposes, and spreads to the systems of the patient's assigned doctors to check for permissions allowing the nurse access to the records. Finally, if the permission is given, the decomposed monitors migrate back to the nurse's device to check that the records are available. As with choreographed monitoring, and in contrast to orchestration, migrating monitors can ensure that monitoring is performed locally. The main difference is that instrumentation of monitors can be performed 
at runtime. For instance, when monitoring the hospital contract clause, no monitor is installed on a doctor's system unless a nurse has made a request for information about a patient assigned to that doctor, which is less intrusive. 
\end{example}


The added expressivity of migrating monitors requires a trust management infrastructure to ensure safe deployment of received monitors. Various solutions can be applied towards this end, from monitors signed by a trusted entity showing that they are the result of an approved contract negotiation procedure, to proof-carrying monitors which come with a proof guaranteeing what resources they access. This issue will not be discussed further here, but is crucial for the practicality of migrating monitors.

There are a number of issues relating to these different monitoring approaches that are unresolved.  For instance, it is somewhat unclear, at least from a formal perspective, what added benefits migration brings to distributed monitoring.  There are also issues relating to the monitoring of consequential properties across locations, which cannot be both sound and complete: in Example \ref{eg:general-example}, by the time the monitor migrates to the doctor's system, the doctor may have already approved the nurse's request. Distribution precludes precise analysis of the relative timing of the traces, and one has the option of taking a worst or best case scenario, avoiding false positives or false negatives respectively.  This problem is also prevalent in both orchestrated and choreographed approaches. 
We therefore require a \emph{common formal framework} where all three approaches can be expressed. This would, in turn, permit rigorous analysis and evaluation with respect to these issues. 




\section{A Distributed Monitoring Language}
\label{sec:language}

We present \MDPi, an adaptation of the distributed $\pi$-calculus in \cite{Hennessy2007}, where processes are partitioned across a flat organisation of  locations; their behaviour is amenable to monitoring through traces, administered at a local level.  The syntax, presented in Figure~\ref{fig:mdpi},  assumes denumerable sets of channel names $c,\,d\in \Chans$, location names $l,k,h\in\Locs$, indices $n,m,o \in \Idx$ 
and variables $x,\,y\in\Vars$; identifiers $u,v$ range over $\Idents\, =\, \Chans\cup\Locs\cup\Idx\cup\Vars$, with identifier list $v_1,\ldots,v_n$ denoted as $\lst{v}$.

\begin{display}[H]{\MDPi\ Syntax}{fig:mdpi}
   \begin{tabular}{rcl}
         \!\!\!$S,R \in \Sys$ & ::=&
          $ \piPLocation{k}{P} ~~|~~ S \piPar R ~~|~~ \piNew{c}{S} $  \\[5pt]
          \!\!\!$P,Q \in \Proc$ & ::=&
          $ \piStop \,~|\,~ \piOut{u}{v}{P} ~~|\,~ \piIn{u}{x}{P} ~~|\,~ \piNew{c}{P} ~~|~~
           \piIfThenElse{u \!=\! v}{\!P}{\!Q} ~~|
          ~~ P\! \piPar\! Q ~~|~ \piRepeat{P}   ~~|\,~
             \piMonitor{M}{l,n} \,~|\,~T$\\[5pt]
            \!\!\!$T \in \Trac$ & ::=&
           $ \piMonTrace{c}{\lst{d}}{n} $\\[5pt]
         \!\!\!$M,N \in \Mon$ & ::=&
          $ \piStop ~~|~ \piOut{u}{v}{M} ~~|~
          \piIn{u}{x}{M} ~~|~~ \piNew{c}{M} ~~|~~
           \piIfThenElse{u \!=\! v}{M}{N} ~~|
          ~~ M \!\piPar\! N ~~|~ \piRepeat{M} $\\[5pt]
           & $|$ & $ \piMonIn{c}{\lst{x}}{M} ~|~~ \piMSync{u}{M}
              ~|~~ \piMGetC{x}{y}{M} ~|~~ \piMSetC{u}{v}{M}
              ~|~~\piGo{u}{M} ~|~~ \piOk ~|~~ \piFail    $
        \end{tabular}
\end{display}

\textit{Systems}, $S,R$,  are made up of \textit{located processes} $\piPLocation{k}{P}$ 
(the tag $k$ denotes the current location hosting $P$
) which can be composed in parallel and subject to scoping of channel names.

Located entities are partitioned into three syntactic categories: \emph{Processes}, \Proc, comprise the standard communication constructs for output,  \piOut{c}{\lst{d}}{P}, and  input, \piIn{c}{\lst{x}}{P} (variables $\lst{x}$ are bound in the continuation $P$), together with the name-matching conditional, replication, parallel composition and name restriction; \emph{Traces}, are made up of individual trace entities, $\piMonTrace{c}{\lst{d}}{n}\in\Trac$ recording communication of values $\lst{d}$ on channel $c$  at timestamp $n$ - they are meant to be ordered as a complete log recording past computation at a particular location; \emph{Monitors}, \Mon, are similar in structure to processes, and are delimited at the process level by enclosing brackets $\piMonitor{M}{k,n}$, where $(k,n)$ denotes the \emph{monitoring context \ie the location and log position of the trace being monitored}. In addition, they can:
\begin{itemize}
\item query traces for records of communication on channel $c$, \piMonIn{c}{\lst{x}}{M}, where the index and location of the trace monitored is inferred from the enclosing  monitoring context, $(k,n)$.
\item get the information relating to the current monitoring context, $\piMGetC{x}{y}{M}$ ($x, y$ bound in $M$), and set the monitoring context to specific values $k$ and $n$, $\piMSetC{k}{n}{M}$, or else update to the current timestamp of a location $k$, $\piMSync{k}{M}$,
\item migrate to location $k$, \piGo{k}{M}, and
\item report success, \piOk, or failure, \piFail.
\end{itemize}

\paragraph*{Shorthand:}
\label{sec:shorthand}
We often elide trailing \piStop\ processes. We thus represent asynchronous outputs such as \piOut{c}{\lst{v}}{\piStop} as \piOutA{c}{\lst{v}} and branches such as \piIfThenElse{u=v}{P}{\piStop} as \piIfThen{u=v}{P}.  We also denote $\piMLocation{k}{M}{l,n}$ as syntactic sugaring  for $\piPLocation{k}{\piMonitor{M}{l,n}}$.

\medskip

Our calculus describes distributed, event-based, asynchronous monitoring. 
Monitoring is \emph{asynchronous} because it happens in two phases, whereby the operational mechanism for tracing is detached from the operational mechanism for querying the trace.  This two-set setup closely reflects the limits imposed by a distributed setting and is more flexible with respect to the various monitoring mechanisms we want to capture.   Monitoring is \emph{event-based} because we chose only to focus on recording and analysing discrete events such as communication.

For simplicity, our calculus only records effected output process events in traces (which dually imply effected inputs); we can however extend traces to record other effects of computation in straightforward fashion.
In order to extract realistic temporal ordering of traces across locations, the calculus provides two mechanisms for monitor re-alignment: the coarser (and real-time) \kw{sync} operation  is used to start monitoring from a particular instant in time;
the more explicit context update \kw{setI}  enables hard-coded control of relative timing at the level of monitors. For instance, together with \kw{getI}, it enables decomposed monitoring to hand over tracing at a specific index in a local trace. This mechanism gives more control and can improve distributed monitoring precision, but may also lead to unsound monitoring. 



\paragraph*{Tracing:}
\label{sec:tracing}

Whenever communication occurs (possibly across two locations) on some channel $c$ with values $\lst{v}$, a trace entity of the form \piMonTrace{c}{\lst{v}}{n} is produced at the location where \emph{the output resides}. The output location, which keeps a local counter, assigns the timestamp $n$ to this trace entity and increments its counter to $n+1$. Local timestamps induce a partial-order amongst all trace entities \textit{across the system}.  In particular, we obtain a finite (totally ordered) chain of traces \textit{per location}.

\begin{example}[Distributed Tracing]\label{eg:distr-tracing}  Consider the distributed system of outputs:
     $$\piPLocation{l}{\piOutA{c_1}{v_1}} \;\piPar\; \piPLocation{l}{\piOutA{c_2}{v_2}} \;\piPar\; \piPLocation{k}{\piOutA{c_3}{v_3}}$$
Assuming that locations $l$ and $k$ have the respective timestamp counters $n$ and $m$, once all outputs are consumed we can obtain either of the following possible sets of trace entities (for simplicity, we assume that $v_1\neq v_2$):
\begin{eqnarray}
  \label{eq:1}
  & \piPLocation{l}{\piMonTrace{c_1}{v_1}{n}}, \; \piPLocation{l}{\piMonTrace{c_2}{v_2}{n+1}},\; \piPLocation{k}{\piMonTrace{c_3}{v_3}{m}}\\
  \label{eq:2}
   &  \piPLocation{l}{\piMonTrace{c_1}{v_1}{n+1}}, \; \piPLocation{l}{\piMonTrace{c_2}{v_2}{n}},\; \piPLocation{k}{\piMonTrace{c_3}{v_3}{m}}
\end{eqnarray}
The timestamps of trace-set (\ref{eq:1}) record the fact that the output on $c_1$ was consumed \emph{before} that on $c_2$, whereas those of  trace-set (\ref{eq:2}) record the opposite.  However, in both of these trace-sets, the timestamp assigned to the trace-entity relating to $c_3$, recorded at location $k$, does not indicate the order it was consumed, relative to the outputs on $c_1$ and $c_2$, which occurred at location $l$.
\end{example}

\paragraph*{Concurrent Monitoring:}
\label{sec:monitoring}
In our model, traces may be queried by multiple concurrent entities, which allows for better separation of concerns when monitors are instrumented.  Trace querying is performed exclusively by monitors, $\piMLocation{l}{M}{k,n}$, parameterised by the monitoring context $(k,n)$; indicating that monitor $M$ is interested in analysing the $n^\text{th}$ trace record at location $k$.


\begin{example}[Parallel Monitoring]\label{eg:conc-monitoring}  Consider the trace-set (\ref{eq:1}) from Example~\ref{eg:distr-tracing}.  A (local) monitor  determining whether, from timestamp $n$ onwards, a value $v_2$ was communicated on the first output on channel $c_2$ at location $l$,  can be expressed as:
  \begin{displaymath}
    \piMLocation{l}{\piMonIn{c_2}{x}{\piIfThenElse{x=v_2}{\piOk}{\piFail}}}{l,n}
  \end{displaymath}


The counter $n$ of this monitor indicates that it starts analysing the trace-set (\ref{eq:1}) from the trace entity  $\piPLocation{l}{\piMonTrace{c_1}{v_1}{n}}$ and continues moving up the chain of trace entities until the first trace entity describing outputs on channel $c_2$ is encountered.  Since $\piPLocation{l}{\piMonTrace{c_1}{v_1}{n}}$ states that the event occurred on another channel, namely  $c_1$, the monitor skips the irrelevant trace entity and its index  is incremented to $n+1$ \ie $\piMLocation{l}{\piMonIn{c_2}{x}{\piIfThenElse{x=v_2}{\piOk}{\piFail}}}{{l,n+1}}$.  The monitor analyses the trace entity with the incremented timestamp $n+1$ \ie $\piPLocation{l}{\piMonTrace{c_2}{v_2}{n+1}}$, which happens to match the required event on channel $c_2$; the monitor thus substitutes $v_2$, obtained from the trace entity  $\piPLocation{l}{\piMonTrace{c_2}{v_2}{n+1}}$, for $x$ and proceeds with the monitor processing, which should eventually yield \piOk.    Traces can be concurrently queried by multiple monitors. For instance, consider another monitor running in parallel with the previous one of the form:
\begin{displaymath}
    \piMLocation{l}{\piMonIn{c_1}{x}{\piMonIn{c_2}{y}{\piIfThenElse{x=y}{\piOk}{\piFail}}}}{l,n}
  \end{displaymath}
which checks that equal values are communicated on channels $c_1$ and $c_2$.   Thus it is important that trace entities, such as $\piPLocation{l}{\piMonTrace{c_2}{v_2}{n+1}}$, are persistent and \emph{not} consumed once analysed, as in the case of outputs in a message passing setting.
\end{example}

\paragraph*{Distributed Monitoring:}
\label{sec:distr-monit}

Distribution adds another dimension of complexity to monitor instrumentation, in terms of how to partition monitors across locations and how this partitioning evolves as computation progresses.  In our calculus, \emph{remote} querying can be syntactically expressed as  $\piMLocation{k}{\piMonIn{c}{x}{M}}{l,n}$ for some $c$ and $n$, where $k\neq l$.  We can also describe the different classifications of distributed monitoring outlined earlier in Section~\ref{sec:distmon}.

 \begin{example}\label{eg:dist-monitors}
Consider a distributed system
\begin{equation*}
  \textsl{Sys} \quad\df\quad  \piPLocation{l}{\piIn{c_1}{x}{\piOutA{c_2}{x}}} \;\piPar\; \piNew{d}{(\;\piPLocation{k}{\piIn{d}{x}{\piOutA{c_2}{x}}} \;\piPar\; \piPLocation{l}{\piOut{c_1}{v}{\piOut{d}{v}{\piIn{c_2}{x}{\piStop}}}}\;)}
\end{equation*}
 where  $\piIn{c_1}{x}{\piOutA{c_2}{x}}$ and $\piOut{c_1}{v}{\piOut{d}{v}{\piIn{c_2}{x}{\piStop}}}$ are located at $l$ whereas $\piIn{d}{x}{\piOutA{c_2}{x}}$ is located at $k$.  Moreover, process $\piIn{d}{x}{\piOutA{c_2}{x}}$ shares a scoped channel  $d$ with $\piOut{c_1}{v}{\piOut{d}{v}{\piIn{c_2}{x}{\piStop}}}$.   For some timestamps $n$ and $m$, \textsl{Sys} non-deterministically produces either of the traces (\ref{eq:3}) or (\ref{eq:4}) below; the non-determinism is caused by the competition for the output on channel $c_2$ by respective inputs at $l$ and $k$:
 \begin{eqnarray}
   \label{eq:3}
   &\piNew{d}{(\;\piPLocation{l}{\piMonTrace{c_1}{v}{n}}, \; \piPLocation{l}{\piMonTrace{d}{v}{n+1}},\; \piPLocation{l}{\piMonTrace{c_2}{v}{n+2}}\;)}  \\
   \label{eq:4}
   & \piNew{d}{(\;\piPLocation{l}{\piMonTrace{c_1}{v}{n}}, \; \piPLocation{l}{\piMonTrace{d}{v}{n+1}},\; \piPLocation{k}{\piMonTrace{c_2}{v}{m}}\;)}
 \end{eqnarray}

The preservation of the property  \textit{`Whenever a value $e$ is communicated  on the first output on $c_1$ at location $l$, then this value is \textbf{not}  output on a subsequent output on channel $c_2$  at $k$'} in general cannot be adequately determined statically, due to the non-deterministic nature of the computation,  as exhibited by the possible traces (\ref{eq:3}) and (\ref{eq:4}). However, the property can be monitored at runtime in a number of ways:

$$\begin{array}{lcl}
M^\text{orch} & \df &
  \piMLocation{h}{
       \piMonIn{c}{x}{\;\piIfThen{x=e}{\piMSync{k}{\;\piMonIn{c}{y}{\;\piIfThen{x=y}{\piFail}}}}
  }}{l,n} \\
 M^\text{chor} & \df  &
   \piNew{d'}{(\;\piMLocation{l}{
   \piMonIn{c}{x}{\;\piIfThen{x=e}{\piOutA{d'}{x}}}}{l,n} \; \piPar\;
      \piMLocation{k}{\piIn{d'}{x}{\;\piMSync{k}{\; \piMonIn{c}{y}{\;\piIfThen{x=y}{\piFail}  }}}}{k,m}
  \;)}
\\
M^\text{mig} & \df &
 \piMLocation{l}{
    \piMonIn{c}{x}{\;\piIfThen{x=e}{\piGo{k}{\piMSync{k}{\piMonIn{c}{y}{\piIfThen{x=y}{\piFail}  }}}}}}{l,n}
\end{array}
$$

$M^\text{orch}$ monitors for this property in \emph{orchestrated} fashion, querying traces at both $l$ and $k$ from a remote central location $h$; this monitor is well-aligned with location $l$ to start with, but has to explicitly re-align with location $k$ once monitoring shifts to that location. $ M^\text{chor}$ is an instance of a \emph{choreographed} monitor setup, instrumenting local monitors at each location where trace querying needs to be performed, namely $l$ and $k$.  These local monitors synchonise between them using remote communication on the scoped channel $d'$. Note that the monitor at $k$ updates its context upon channel synchronisation on $d'$ to ensure a temporal ordering on analysed trace records; without synchronisation, the monitor would potentially be reading past parts of the trace which may lead to unsound sequentiality conclusions. Finally, $M^\text{mig}$ is a case of a \emph{migrating monitor}, that starts monitoring at location $l$ but then migrates to location $k$ if it needs to continue monitoring there, re-aligning its index to that of the destination location.

\end{example}

All three distributed monitors in Example~\ref{eg:dist-monitors} are sound \wrt the property stated, in the sense that they never falsely flag a violation.  They are nevertheless incomplete, and may miss out on detecting property violations.  For instance,  $M^\text{orch}$  may realign with location $k$ after the trace $\piPLocation{k}{\piMonTrace{c_2}{v}{m}}$ is generated by $k$, which sets the monitor timestamp index to $(k,m+1)$.  This forces the monitoring to start querying the trace at $k$ from index $m+1$ and will therefore skip the relevant trace item $\piPLocation{k}{\piMonTrace{c_2}{v}{m}}$.  This aspect is however not a limitation of our encoding, but rather an inherent characteristic of distributed computing as discussed earlier in Section~\ref{sec:distmon}.

\section{Monitoring Semantics}
\label{sec:monitoring-semantics}

We define the semantics of \MDPi\ in terms of a number of related Labelled Transition Systems (LTSs), which are then used to compare systems through the standard notion of weak-bisimulation equivalence, denoted here as $\approx$.   This framework allows us to state and prove properties from a behavioural perspective about our monitored systems.  For instance,  we could express the fact that, ignoring monitoring location, $M^\text{orch}$ and $M^\text{chor}$ from Example~\ref{eg:dist-monitors} 
monitor for the same properties \wrt \textsl{Sys}, using the statement:
\begin{equation}
  \label{eq:6}
  \textsl{Sys} \piPar M^\text{orch}\;\approx\; \textsl{Sys} \piPar M^\text{chor}
\end{equation}
Using an LTS that does not express observable monitor actions, the property that a monitor, say  $M^\text{orch}$, does not affect the observable behaviour of the  system \textsl{Sys} could be stated as:
\begin{equation}
  \label{eq:5}
  \textsl{Sys} \;\approx\; \textsl{Sys} \piPar M^\text{orch}
\end{equation}
 Using  different LTSs, the same system could be assigned more restricted behaviour.  For instance, this is useful to ensure that the monitor  $M^\text{mig}$ of Example~\ref{eg:dist-monitors} does not perform remote querying at any stage during its computation by establishing the comparison:
 \begin{equation}
   \label{eq:7}
   \textsl{Sys} \piPar M^\text{mig}\;\approx\; ((\textsl{Sys} \piPar M^\text{mig}) \text{ without remote monitoring})
 \end{equation}
where the lefthand system is subject to an LTS allowing remote querying whereas the righthand monitor is subject to an LTS that prohibits it. Intuitively, if the behaviour is preserved when certain internal moves are prohibited, this means that these moves are not used (in any useful way) by the monitor.

\subsection{Deriving LTSs Modularly}
\label{sec:deriv-ltss-modul}

Closer inspection of  the comparisons (\ref{eq:6}),~(\ref{eq:5}) and (\ref{eq:7}) reveals that the different LTSs required are still expected to have substantial common structure; typically they would differ with respect to either the information carried by actions and/or the type of actions permitted.  For instance, in (\ref{eq:6}) we would want actions that \emph{restrict} information relating to the location of where monitoring is carried out, as this additional information would distinguish between the two monitors.  On the other hand, for (\ref{eq:7}) we would want to \emph{prohibit}  actions relating to remote monitoring.

We therefore construct these related LTSs in modular fashion through the use of a \emph{preLTS}, \ie an LTS whose transitions relate more systems, and whose action labels carry more information than actually needed.  The excess transitions and label information are then pruned out as needed by a \emph{filter function} from actions in the preLTS to actions in the LTS required.

\subsection{A preLTS for \MDPi}
\label{sec:pre-lts-mdpi}

Our preLTS is defined over systems subject to a \emph{local logical clock} at every location used by the system, which are used to generate ordered trace-entities and to re-align monitors.  These clocks are modelled as monotonically increasing counters and expressed as a  partial function $\delta\in\Delta :: \Locs \rightharpoonup \Nat$, where $\delta(l)$ denotes the next timestamp to be assigned for a trace entity generated at $l$. Moreover, the counter increment is defined using standard function overriding, $\inc{\delta}{k} = \delta[k\mapsto (\delta(k)+1)]$.

\begin{display}[h]{\MDPi\ preLTS main rules}{fig:mdpi-preLTS}
    \begin{mathpar}
\inferrule*[left = \rtit{OutP}]
{  }
{
 \conf{\delta}{\piPLocation{k}{ \piOut{c}{d}{P} } }
     \piTrPad{ \loutP{c}{\lst{d}}{k,l}}
 \conf{\kw{inc}(\delta, k)}{\piPLocation{k}{ P } \piPar \piPLocation{k}{\piMonTrace{c}{\lst{d}}{\delta(k)}}}
}
\quad\; \inferrule*[left = \rtit{InP}]
{  }
{
 \conf{\delta}{\piPLocation{l}{ \piIn{c}{x}{P} }}
     \piTrPad{ \linP{c}{\lst{d}}{k,l}}
 \conf{\delta}{\piPLocation{l}{ P \subC{\lst{d}}{\lst{x}}} }
}

\inferrule*[left = \rtit{OutT}]
{  }
{ \conf{ \delta }{\piPLocation{k}{ \piMonTrace{c}{\lst{d}}{n} }}
   \piTrPad{ \tr{k,l}{c}{\lst{d}}{n}}
   \conf{\delta}{\piPLocation{k}{ \piMonTrace{c}{\lst{d}}{n} }}
}
\quad\;
\inferrule*[left = \rtit{InT}]
{  }
{ \conf{ \delta }{\piMLocation{l}{ \piMonIn{c}{\lst{x}}{M} }{k,n}}
 \piTrPad{ \mn{l}{c}{\lst{d}}{k}{n}}
   \conf{\delta}{\piMLocation{l}{ M \subC{\lst{d}}{\lst{x}} }{{k,n+1}}}
}

\inferrule*[left = \rtit{OutM}]
{\vspace{1cm}  }
{ \conf{ \delta }{\piMLocation{k}{ \piOut{c}{d}{M} }{l,n}}
   \piTrPad{ \loutM{c}{\lst{d}}{k,h}}
   \conf{\delta}{\piMLocation{k}{ M }{l,n}}
}
\qquad\;
\inferrule*[left = \rtit{InM}]
{  }
{
 \conf{\delta}{\piMLocation{l}{ \piIn{c}{x}{M} }{k,n}}
     \piTrPad{ \linM{c}{\lst{d}}{h,l}}
 \conf{\delta}{\piMLocation{l}{M \subC{\lst{d}}{\lst{x}}}{k,n}}
}

\inferrule*[left = \rtit{Open}, right = {$[b\neq c, b\in\lst{d}]$} ]
{  \conf{\delta}{S}
     ~\piTrPad{ \loutgen{(\lst{b})c}{\lst{d}}}~
 \conf{\delta'}{S'}}
{
 \conf{\delta}{\piNew{b}{S}}
     ~\piTrPad{ \loutgen{(b,\lst{b})c}{\lst{d}}}~
 \conf{\delta'}{S'}
}

\inferrule*[left = \rtit{Res}, right ={$[b \not{\in} \fn{\mu}]$}]
{ \conf{\delta}{S}
     ~\piTrPad{ \mu }~
 \conf{\delta'}{S'} }
{
  \conf{\delta}{\piNew{b}{S}}
     ~\piTrPad{ \mu }~
 \conf{\delta'}{\piNew{b}{S'}}
}

\inferrule*[left = \rtit{Com1}, right = {$[\lst{b}\cap\fn{R} = \emptyset]$}]
{  \quad\conf{\delta}{S}
     ~\piTrPad{ \loutgen{(\lst{b})c}{\lst{d}}}~
 \conf{\delta'}{S'}   \qquad \qquad \conf{\delta}{R}
     ~\piTrPad{ \lingen{c}{\lst{d}}}~
 \conf{\delta}{R'} \quad}
{
 {\conf{\delta}{S \piPar R}
     ~\piTrPad{ \ltaugen}~ \conf{\delta'}{\piNew{\lst{b}} (S'\piPar R')}}}

\inferrule*[left = \rtit{Skip}, right = {$[c_1 \neq c_2]$}]
{\quad  \conf{\delta}{S}
     ~\piTrPad{ (\lst{b})\tr{l,k}{c_1}{\lst{d}}{n} }~
 \conf{\delta}{S}   \qquad \qquad
 \conf{\delta}{\piMLocation{k}{ M }{{l,n}}}
     ~\piTrPad{ \mn{k}{c_2}{\lst{e}}{l}{n}}~
 \conf{\delta}{\piMLocation{k}{ M' }{{l,n+1}}} \quad}
{\conf{\delta}{S \piPar \piMLocation{k}{ M }{{l,n}}}
   ~\piTrPad{\ltauT{l}{k}{l,n}}~
   \conf{\delta}{S \piPar \piMLocation{k}{ M }{{l,n+1}}}
}

\inferrule*[left = \rtit{SetI}]
{ }
{
 \conf{\delta}{\piMLocation{k}{ \piMSetC{h}{m}{M} }{l,n}}
     \piTrPad{ \ltauM{k}{k} }
 \conf{\delta}{\piMLocation{k}{M}{h,m}}
}
\quad
\inferrule*[left = \rtit{Sync}]
{  }
{
 \conf{\delta}{\piMLocation{k}{ \piMSync{l}{M} }{h,n}}
     \piTrPad{ \ltauM{k}{k} }
 \conf{\delta}{\piMLocation{k}{ M }{{l,\delta(l)}}}
}

\inferrule*[left = \rtit{GetI}]
{ }
{
 \conf{\delta}{\piMLocation{k}{ \piMGetC{x}{y}{M} }{l,n}}
     \piTrPad{\ltauM{k}{k}}
 \conf{\delta}{\piMLocation{k}{M \subC{l,n}{x,y}}{l,n}}
}
\quad
\inferrule*[left = \rtit{Go}]
{  }
{
  \conf{\delta}{\piMLocation{k}{ \piGo{l}{M} }{h,n}}
       \piTrPad{\ltauM{k}{l}}
  \conf{\delta}{\piMLocation{l}{ M }{{h,n}}}
}
\end{mathpar}
\end{display}

A \emph{Configuration} $C,D \in \Conf :: \Delta \times \Sys$ is thus a system subject to a set of localised counters. The preLTS is a ternary relation $\rightarrow :: \Conf\times\pAct\times\Conf$, denoted using the suggestive notation $C \piTr{\mu} D$, where $\mu \in \pAct$ is a preLTS action label of the form, $\ltaugen$, an internal action, $(\lst{b})\loutgen{c}{\lst{d}}$, an output action, or $\lingen{c}{\lst{d}}$, an input action. In case of the output action, $\lst{b}$ denotes the (possibly empty) set of channel names exported during an eventual interaction. These actions are standard \cite{Hennessy2007}, but are decorated with additional information \actinf\ which can be of the following three formats:
\begin{description}
\item[$\langle p:l,k \rangle -$] This states that it is a \emph{process} ($p$) action, involving locations $l$ and $k$.
\item[$\langle m:l,k \rangle -$] This states that it is a \emph{monitor} ($m$) action, involving locations $l$ and $k$.
\item[$\langle t:l,k:n \rangle -$] This states that it is a \emph{trace} ($t$) action at timestamp $n$, involving locations $l$ and $k$, .
\end{description}

The main rules defining the relation $C \piTr{\mu} D$ are outlined in Figure~\ref{fig:mdpi-preLTS}.  The rule for process input, \rtit{InP}, is standard, except for the additional label tag $\langle p:k,l \rangle$ encoding the fact that the input is a process input, it resides at location $l$, and is reading from some location $k$ (when communication is local, then $l=k$).  A central rule to our monitoring semantics is \rtit{OutP}. Apart from the additional label decoration, it differs from standard output rules in two respects: first it generates a  trace entity, $\piPLocation{k}{\piMonTrace{c}{\lst{d}}{\delta(k)}}$, recording the channel name, $c$, the values communicated, $d$, timestamped by $\delta(k)$, and second, it increments the clock at $k$ once the trace entity is generated, necessary for generating a total order of trace-entities at $k$.  Monitor communication, defined by rules \rtit{OutM} and \rtit{InM}, is similar albeit simpler since neither trace entities are generated, nor is the local counter updated\footnote{Note that rule \rtit{OutM} refers to an indeterminate location $h$, to match a reader in any such location.}.

Rule \rtit{OutT} models trace actions as output labels with tags $\langle t:k,l:n \rangle$, where the timestamp of the trace, $n$, is recorded in the tag as well.  Crucially, the trace entity is not consumed by the action (thereby acting as a broadcast), and its persistence allows for multiple monitors to query it.  This action can be matched by a query action, \rtit{InT}, expressed as an input action with a matching tag  $\langle t:k,l:n \rangle$ where the source location of the trace entity, $k$, and time stamp $n$ must match the current monitoring context $(k,n)$.  Since the action describes the fact that a trace entity has been matched by the monitor query, the timestamp index of the monitoring context is incremented, $(k,n+1)$ to progress to the next entitity in the local trace log.

Scope extrusion of channel names may occur both directly, through process or monitor communication, or else indirectly through trace querying; these are both handled  by the standard scoping rules \rtit{Open} and \rtit{Res}.  All three forms of communication \ie process, monitor and trace, are also handled uniformly, this time by the communication rule \rtit{Com1}  (we here elide its symmetric rule).  Communication yields a silent action $\ltaugen$ that is decorated with the corresponding tagging information from the constituent input and output actions of the premises.  This tagged information must match for both input and output actions and, in the case of the trace tags, $\langle t:k,l:n \rangle$, this also implies a matching of the timestamp $n$.  When, for a particular timestamp, querying does not match the channel of the trace entity at that timestamp, rule \rtit{Skip} allows the monitor to increase its timestamp index and thus querying to move up the trace-chain at that location.  Finally, \rtit{Sync} allows monitors to realign with a trace at a particular location, \rtit{GetI} and \rtit{SetI} allow for explicit manipulations of the monitoring context whereas \rtit{Go} describes monitor migration. 

\subsection{Filter Functions}
\label{sec:filter-functions}

Although necessary to encode extended information of system execution, the preLTS presented is too discriminating.  For instance, the internal action $\ltaugen$ is now compartementalised into distinct silent actions, each identified by the tag information $\actinf$, which complicates their use for weak actions when verifying bisimilar configurations.  Similarly, external actions differentiating between a process or a monitor carrying out that action may also be deemed to discriminating.  Finally, we may also want to disallow certain actions such as remote trace querying.

We obtain LTSs with the necessary level of discriminating actions using (i) the preLTS of Section~\ref{sec:pre-lts-mdpi}, together with (i) a filter function, $\Omega$.  This function maps actions in the preLTS, $\mu \in \pAct$, to actions in the required LTS, $\alpha \in \Act$,  through the rule:
\begin{mathpar}
\inferrule*[left = \rtit{Fltr},right = {$[\Omega(\mu) = \alpha]$}]
{ C_1 \piTrPad{\mu} C_2  }
{C_1 \piTrPad{\alpha}_\Omega C_2 }
\end{mathpar}
\paragraph*{Notation:}
\label{sec:notation}
Note that filter function applications are essentially abstractions of the preLTS. LTSs obtained  in this manner can effectively be indexed by their respective filter function, $\Omega$, and for clarity  we denote a configuration $C$ subject to a behaviour obtained from the preLTS and a filter function $\Omega$ as $C_\Omega$.  We also denote transitions obtained in this form as $C_1 \piTrPad{\alpha}_\Omega C_2 $.   

\begin{example}[Filter Functions]\label{eg:filter-functions}  Consider the following filter function definitions:
\[
  \begin{array}{lcllcllcl}
    \Omega_\textsl{NTg}(\ltaugen)        & \df & \tau    &
    \Omega_\textsl{Prc}(\ltaugen)        & \df & \tau    &
    \Omega_\textsl{LTr}(\ltauT{l}{l}{n}) & \df &  \tau\\

    \Omega_\textsl{NTg}(\loutgen{c}{\lst{d}})      & \df & c ! \lst{d}              \qquad\qquad&
    \Omega_\textsl{Prc}(\lout{c}{\lst{d}}{l,k}{p}) & \df & \loutNL{c}{\lst{d}}{l,k} \qquad\qquad&
    \Omega_\textsl{LTr}(\ltau{l}{k}{\_})           & \df & \tau  \\

    \Omega_\textsl{NTg}(\lingen{c}{\lst{d}})      & \df & c ? \lst{d} &
    \Omega_\textsl{Prc}(\lin{c}{\lst{d}}{l,k}{p}) & \df & \linNL{c}{\lst{d}}{l,k} &
    \Omega_\textsl{LTr}(\loutgen{c}{\lst{d}})     & \df & c ! \lst{d}\\

    &&&
    &&&
    \Omega_\textsl{LTr} (\lingen{c}{\lst{d}}) & \df & c ? \lst{d}
  \end{array}
\]

  $\Omega_\textsl{NTg}$ removes all tags from decorated  actions, which in turn allows for a straightforward definition of weak actions, $\piTrWPad{\hat{\alpha}}$, as $(\piTrPad{\tau})^\ast$ if $\alpha=\tau$ and $(\piTrPad{\tau})^\ast \piTrPad{\alpha} (\piTrPad{\tau})^\ast$ otherwise.   In addition to stripping $\tau$ action tags, the second filter function, $\Omega_\textsl{Prc}$, allows only process external actions, filtering out the $p$ component in this case.   The function is partial as it is undefined for all other preLTS actions. This is useful when we do not want to discriminate configurations based on the tracing and monitoring actions.  The final filter function,
$\Omega_\textsl{LTr}$, removes all tags but uses them to prohibit silent tracing actions where the two locations in the tag are distinct; this in effect rules out remote trace querying, thereby enforcing \emph{localised trace monitoring}.
\end{example}

We assume a certain well-formedness criteria on our filter functions, such as that they do not change the form of an action (\eg an output action remains an output action), and that whenever they map to  silent actions, $\tau$, these are not decorated; the filter functions in Example~\ref{eg:filter-functions} satisfy these criteria. Through this latter requirement we re-obtain the standard silent $\tau$-action at the LTS level.

\subsection{Behavioural Equivalence}
\label{sec:behav-equiv}


The technical development in sections~\ref{sec:pre-lts-mdpi} and \ref{sec:filter-functions} allows us to immediately apply weak bisimulation \cite{Hennessy2007} as a coinductive proof technique for equivalence between LTSs obtained for our preLTS and well-formed filter functions.  Two (filtered) LTSs, $C_{\Omega_1}$ and $D_{\Omega_2}$, are bisimilar, denoted as $C_{\Omega_1} \bisim D_{\Omega_2}$, if they match each other's transitions; we use the weak bisimulation variant, $\bisim$, as this abstracts over internal $\tau$-actions which yields a more natural extensional equivalence.


\begin{example} \label{eg:bisim-statements}
  Using the filter functions defined in
  Example~\ref{eg:filter-functions}, we can formally state and prove
  equivalences (\ref{eq:6}), (\ref{eq:5}) and (\ref{eq:7}) outlined
  earlier, for a localised clock-set $\delta$ including
  locations $l$ and $k$:
\begin{eqnarray}
 \label{eq:1sem}
  (\conf{\delta}{\textsl{Sys} \piPar M^\text{orch}})_{\Omega_\textsl{NTg}}  &\quad \bisim \quad & (\conf{\delta}{\textsl{Sys} \piPar M^\text{chor}})_{\Omega_\textsl{NTg}} \\
  \label{eq:2sem}
  (\conf{\delta}{\textsl{Sys}})_{\Omega_\textsl{Prc}}  & \quad \bisim \quad & (\conf{\delta}{  \textsl{Sys} \piPar   M^\text{orch}})_{\Omega_\textsl{Prc}}\\
  \label{eq:3sem}
  (\conf{\delta}{ \textsl{Sys} \piPar M^\text{mig}})_{\Omega_\textsl{NTg}} & \quad \bisim \quad & (\conf{\delta}{ \textsl{Sys} \piPar M^\text{mig}})_{\Omega_\textsl{LTr}}
\end{eqnarray}
Equivalence (\ref{eq:1sem}) formalises the behaviour expected for (\ref{eq:6}) using an LTS whose actions prohibit distinctions based on action tags; including monitoring location, \ie  $\Omega_\textsl{NTg}$.   Since in (\ref{eq:5}) we wanted to analyse how monitors affect process computation, in its corresponding equivalence (\ref{eq:2sem})  we use an LTS that tags process external actions with location information while prohibiting any actions relating to tracing or monitoring.  Finally (\ref{eq:3sem}) compares $M_\text{mig}$ with itself, subject to a restricted semantics where remote monitoring is prohibited, \ie $\Omega_\textsl{LTr}$.
\end{example}

\newcommand{\re}[2]{(#1)@#2}
\newcommand{\reA}[2]{#1@#2}

\section{Instrumentation for Distributed Monitoring}
\label{sec:regexp}

The instrumentation of contracts as distributed monitors is non-trivial and can easily lead to unsound contract monitoring. In this section we illustrate how the instrumentation of contracts, expressed using a simple regular expression-based temporal logic specifying violation traces, can be safely automated according to different monitoring approaches. 
The syntax of the contract language is:
\begin{eqnarray*}
E & ::= & \re{c,\lst{v}}{k} \quad|\quad E.E \quad|\quad E^* \quad|\quad E+E
\end{eqnarray*}

\emph{Basic events} have the form $\re{c,\lst{v}}{k}$ indicating that a communication on channel $c$ with value $\lst{v}$ occurs at location $k$. We adopt a semantics allowing for multiple matches, rather than opt only for the shortest match\footnote{In any case, when runtime monitoring one may choose to halt the system on the shortest match.} and thus any trace terminating with a communication $\piOutA{c}{\lst{v}}$ at location $k$ is considered to be a violating trace. The other operators are the standard ones used in regular expressions:  $E.F$ corresponds to the traces which can be split into two, with the first matching $E$, and the second matching $F$; expression $E^*$ corresponds to traces which can be split into a number (possibly zero) parts, each of which satisfies $E$; and  $E+F$ corresponds to the set of traces which match either $E$ or $F$.
\paragraph*{Notation:}
\label{sec:notation-re}
$\sum_{e\in I} E$ corresponds to the generalised choice over finite $I$, which is equal to $E\subC{i_1}{e}+E\subC{i_2}{e}+\ldots E\subC{i_n}{e}$ (where $I=\{i_1,i_2\ldots, i_n\}$).

Despite the apparent simplicity of this expository contract language, we can already express interesting contracts.

\begin{example} Consider a simplification of the contract outlined in Example~\ref{eg:general-example}: \emph{``The release of a patient's
    record must be approved by supervising doctors.''} Stated in terms
  of what leads to a violation, we get: \emph{``If a patient's
    medical record is released regardless of a doctor's disapproval,
    the contract is violated''} which can be expressed as the regular
  expression:
  $$ \sum_{p\in \mtxt{Patient}} \re{\textsf{req},()}{p} .
    \sum_{d\in\mtxt{Doctor}} \re{\textsf{withhold},p}{d} \;.\;\;
    \re{\textsf{send},p}{h} $$
  where  $p$, $d$ and $h$ are locations referring to the patient's,
  the doctor's and the hospital domain, channel names $\textsf{req}$,
  $\textsf{withhold}$ and $\textsf{send}$ denote actions requesting,
  withholding and sending medical records, and  sets $\mtxt{Patient}$
  and $\mtxt{Doctor}$ range over the finite patients and doctors in the
  system.
\end{example}

There are different ways in which one may transform a regular expression into an \MDPi\ term. For instance,  $\re{c_1,\overline{v_1}}{k_1}.\re{c_2,\overline{v_2}}{k_2}$  may be matched by either  one monitoring process, $M_1$, or by the split  monitors, $M_2$,  below:
\begin{eqnarray*}
M_1 & \df &
\piMSync{k_1}{
\piMonIn{c_1}{\overline{x_1}}{
  \piIfThen{\overline{x_1}=\overline{v_1}}{
    (\piMSync{k_2}{
       \piMonIn{c_2}{\overline{x_2}}{
          \piIfThen{\overline{x_2}=\overline{v_2}}{\piFail}
       }
    })
  }
}}\\
M_2 & \df &
(
\piMSync{k_1}{
\piMonIn{c_1}{\overline{x_1}}{
\piIfThen{\overline{x_1}=\overline{v_1}}{m !}
}}
) \piPar  
(
\piInGeneric{m}{
\piMSync{k_2}{
  \piMonIn{c_2}{\overline{x_2}}{
    \piIfThen{\overline{x_2}=\overline{v_2}}{\piFail}
  }
}}
)
\end{eqnarray*}


We find that the second translation, $M_2$, lends itself better towards illustrating how monitors can be distributed in different ways across locations --- for example, in an orchestrated approach, we would place all the monitoring processes in a single location, while in a choreographed approach, we would distribute the processes as required. A sequential approach such as $M_1$, may be more approriate in an orchestrated approach (since it avoids unnecessary parallelism), but would not be possible to distribute to enable choreographed monitoring without further manipulation.

\begin{figure}[t]
  \centering
      \resizebox{100pt}{!} {\includegraphics{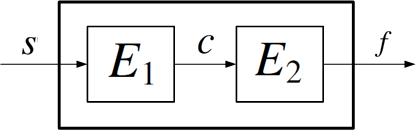}} \hspace{5mm}
      \resizebox{120pt}{!} {\includegraphics{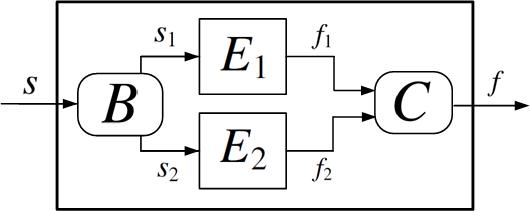}} \hspace{5mm}
      \resizebox{140pt}{!} {\includegraphics{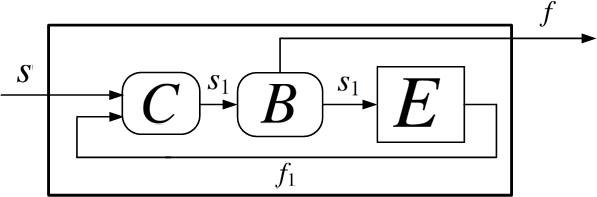}}
\caption{
Compiling $E.F$, $E+F$ and $E^*$ (respectively) where $C, B$ correspond to $\mtxt{comb},\mtxt{bifurc}$.
}
\label{fig:blockdiagrams}
\end{figure}

In this paper, we adopt the maximally parallelised approach, primarily to be able to observe similarities and distinctions between different compilation approaches. In particular we use this translation for a compilation strategy  corresponding closely to standard approaches used in hardware compilation of regular expressions \cite{pace:dcc2002}, producing circuits with two additional wires: an input which is signaled upon to start matching the regular expression, and an output wire which the circuit uses to signal a match with the regular expression. In our case, the wires correspond to channels: basic events $\re{c,\overline{v}}{k}$ with start channel $s$ and match channel $f$ would be translated into an expression which waits for input on channel $s$, then outputs on channel $f$ when an instance of $c$ with $\overline{v}$ occurs at location $k$. Our translations also employ two standard monitor organisations for funneling two output signals into one and forking channel communication  onto two separate ones; these are expressed below as the macros 
$\mtxt{comb}$ and $\mtxt{bifurc}$:
\begin{displaymath}
  \mtxt{comb}
 (f_1,\;f_2,\;f)  \df
  \piRepeat{\,(\,\piIn{f_1}{\lst{x}}{\piOutA{f}{\lst{x}}
    }\,)} \piPar
  \piRepeat{\,(\,\piIn{f_2}{\lst{x}}{\piOutA{f}{\lst{x}}
    }\,)} \qquad\qquad
\mtxt{bifurc}(s,\;s_1,\;s_2)  \df
  \piRepeat{
     \,\piIn{s}{\lst{x}}{ (
        \piOutA{s_1}{\lst{x}}
        \piPar
        \piOutA{s_2}{\lst{x}}
     )}
  }
 \end{displaymath}
We define three compilation strategies,
$\psi_O$, $\psi_C$ and $\psi_{M}$, corresponding respectively to
monitoring using orchestration,
static choreography and migrating monitoring as discussed in section \ref{sec:distmon}.  The compilation procedures use three parameters: two control channels (used to notify when the regular expression is to start being matched, and to notify when it has matched) and the expression to be compiled.  For simplicity, all three translations follow a similar pattern shown by the block diagrams in Figure \ref{fig:blockdiagrams}, varying only in the location placement of the monitors and synchronisation strategy.

\subsection{Orchestration-Based Monitoring Translation}

Orchestration places the monitor at some predefined central location, $h$.  As stated earlier, the lack of a global clock prevents it from deducing with certainty the order of events happening across different locations.  Nevertheless, our translation attempts to mitigate this imprecision for sequence of events occurring \emph{at the same location} using the following mechanism: when a basic event is matched, the monitoring context, $(k,n)$, at that moment is recorded using \kw{getI} and passed as arguments on the match signaling channel; this allows subsequent matching to explicitly adjust the monitoring context to these values using \kw{setI} in cases where the location of subsequent events does not change; in cases where the location changes, this information is redundant and alignment 
is carried out using the coarser \kw{sync} command.
\[ \small
\psi'_O(s,\,f,\,(\re{c,\lst{v}}{k}))  \;\df\;
  \piRepeat{(
    \piIn{s}{(x_\text{loc},x_\text{idx})}{
      \piIfThenElse{k=x_\text{loc}}{
        (\piMSetC{x_\text{loc}}{x_\text{idx}}{\mtxt{trg}(c,\lst{v},f)})
      }{
        (\piMSync{k}{\mtxt{trg}(c,\lst{v},f)})
      }
    }
  )}
\]
In either case, a listener triggers a signal with the updated index of the trace on the match channel for every event $c$ with data $\lst{v}$.  Specifically,  the macro $\mtxt{trg}(c,\lst{v},f)$ repeatedly reads from channel $c$, outputting the monitoring-context information on the matching channel $f$  every time the data traced matches $\lst{v}$:
\[ 
\mtxt{trg}(c,\lst{v},f) \df
            \piRepeat{\,(
                \piMonIn{c}{\lst{x}}{
                  \piIfThen{\lst{x}=\lst{v}}{
                    \piMGetC{x_\text{loc}}{x_\text{idx}}{
                      \piOutA{f}{(x_\text{loc},x_\text{idx})}
                    }
                  }
                }
            )}
\]
The 
compilation of the regular expression operators matches 
the compilation schemata of figure \ref{fig:blockdiagrams}:
\begin{eqnarray*}
\psi'_O(s,\,f,\,(E.F)) & \df &
  \piNew{m}{(\;
    \psi'_O(s,\,m,\,E) \piPar
    \psi'_O(m,\,f,\, F)
  )}\\
\psi'_O(s,\,f,\,E^*) & \df &
  \piNew{c,s',f'}{\;
    \mtxt{comb}(s,f',c) \piPar
    \mtxt{bifurc}(c,s',f) \piPar
    \psi'_O(s',\,f',\,E)
  }\\
\psi'_O(s,\,f,\,(E+F)) & \df &
  \piNew{s_1,s_2,f_1,f_2}{\left(
    \mtxt{bifurc}(s,s_1,s_2) \piPar
    \psi'_O(s_1,\,f_1,\,E) \piPar
    \psi'_O(s_2,\,f_2,\,F) \piPar
    \mtxt{comb}(f_1,f_2,f)\right)}
\end{eqnarray*}
The combined monitors are located at the predefined central location $h$, with a dummy initial monitor context continuation parameters $(h,1)$. The monitor induced for a contract $E$ is thus:
\[
\psi_O(E) \df
  \piMLocation{h}{\;
    \piNew{s,f}{(\;
      \piOutA{s}{(h,1)}  \quad\piPar\quad
      \psi'_O(s,\,f,\,E) \quad\piPar\quad
      \piIn{f}{\lst{x}}{\piFail}
    \;)}
  \;}{h,1}
\]

\subsection{Choreography-Based Monitoring Translation}
 Instead of instrumenting the whole monitor at a single central location, a choreography-based approach decomposes the monitor into parts, possibly placing them at different locations.  Once again, monitors are made up of two kinds of components: (i) the event listeners; and (ii) the choreography control logic made up of $\mtxt{comb}$ and $\mtxt{bifurc}$ components. The event listeners are located locally, where the event takes place, but are otherwise exactly the same as in the orchestrated approach:
\[
\begin{array}{l}
\psi'_C(s,\,f,\,(\re{c,\lst{v}}{k})) \df  \piMLocation{k}{\psi'_O(s,\,f\,(\re{c,\lst{v}}{k}))}{k,1}
\end{array}
\]
On the other hand, the choreography control logic can be placed at any location. For instance one may choose to locate them at the node where the next input will be expected, or where the last one occurred. For a particular choice of locations $l$ and $h$, choice $E+F$ is compiled as follows:
\[
\piNew{s_1,s_2,f_1,f_2}{\left(
     \piMLocation{l}{\mtxt{bifurc}(s,s_1,s_2)}{l,1} \piPar
    \psi'_C(s_1,f_1,E) \piPar
    \psi'_C((s_2,f_2),F) \piPar
    \piMLocation{h}{\mtxt{comb}(f_1,f_2,f)}{h,1}\right)}
\]
Finally, we add the necessary start signal (from some start location $k$) to initiate the monitoring:
\[
\psi_C(E) \df
  \piNew{s,f}{\;
    \piMLocation{k}{
      \piOutA{s}{(k,1)} \piPar
      \piIn{f}{\lst{x}}{\piFail}
    }{k,1} \piPar
    \psi'_C(s,\,f,\,E)
  }
\]

Note that unless all the locations enable the execution of new (monitoring) process at runtime, the contracts must be known at compile-time, which is guaranteed in the simple regular expression logic we are using.

\subsection{Migrating Monitors Translation}
For the migrating monitors technique, we use a simplified translation where the monitors generated are similar to the ones used in orchestration, except that the monitor migrates when required to the relevant location (using the \textsf{go} operator). $\psi'_M$ is defined identical to $\psi'_O$ except for basic events:



\[ \small
\psi'_M(s,f,(\re{c,\lst{v}}{k}))  \;\df\;
  \piRepeat{(
    \piIn{s}{(x_\text{loc},x_\text{idx})}{
      \piGo{k}{
      \piIfThenElse{k=x_\text{loc}}{
        (\piMSetC{x_\text{loc}}{x_\text{idx}}{\mtxt{trg}(c,\lst{v},f)})
      }{
        \piMSync{k}{\mtxt{trg}(c,\lst{v},f)}
      }}
    }
  )}
\]


Note how migration (thus monitor instrumentation) is delayed and happens only once the start signal on channel $s$ is received.
Initially, the monitor can be chosen to reside anywhere. For a particular location choice $h$, the migrating monitor approach for a contract $E$ would be the following:
$$\psi_M(E) \df
  \piMLocation{h}{\;
    \piNew{s,f}{(\;
      \piOutA{s}{(h,1)} \quad\piPar\quad
      \psi'_M(s,\,f,\,E) \quad\piPar\quad
      \piIn{f}{\lst{x}}{\piFail}
    \;)}
  \;}{h,1}
$$
Despite the resemblances resulting from our simplistic translations, migration improves on an orchestrated approach by avoiding remote tracing. As in the choreographed approach, one can also choose to explicitly run the combining and bifurcation processes at particular locations by adding explicit migration instructions. A better approach would be to nest all the monitors within each other to avoid monitors migrating or installed before they are actually required. For example, monitoring for an expression of the form: $\reA{(c_1,\overline{v_1})}{l}.\,\reA{(c_2,\overline{v_2})}{k}.\,\reA{(c_3,\overline{v_3})}{h}$ would be transformed into a monitor of the form:
$$\piGo{l}{\left(\piMonIn{c_1}{\overline{x_1}}{\piIfThen{\overline{x_1} = \overline{v_1}}{
    \piGo{k}{\left(\piMonIn{c_2}{\overline{x_2}}{\piIfThen{\overline{x_2} = \overline{v_2}}{
        \piGo{h}{\left(\piMonIn{c_3}{\overline{x_3}}{\piIfThen{\overline{x_3} = \overline{v_3}}{\piFail}
      }}
    }\right)}
  }\right)}
}\right)}$$ Note that using this approach entails minimal local monitor instrumentation since this happens on a by-need basis: the translation avoids installing any monitor at location $k$ unless $\piOutA{c_1}{\overline{v_1}}$ happens at  $l$.

Even within this simplistic formal setting, migrating monitors can be seen to be more versatile than a choreographed approach.  For instance,   if our 
contract language is extended with variables and a binding construct, $\exists x. E$, we could express a more dynamic form of contract such as $\exists x. \re{c_1,x}{k}.\re{c_2,v}{x}$; in such a contract the location of the second event \emph{depends} on the location communicated in the first event and, more importantly, this location is not known at compile time.  Because of this last point, this contract cannot be handled adequately by traditional choreographed approaches which would need to preemptively instrument monitors at \emph{every location}.  
  However, in a migrating monitor approach, this naturally translates to a single runtime migration.


\subsection{The Approaches and Limitations}


We have shown how one can formulate different monitoring strategies of the same contract using \MDPi.  The contract language and its compilation procedure have  intentionally been kept simple to avoid their complexity from obscuring the underlying monitoring choices. The different approaches mostly differ only in the location of the monitors. The migrating monitor approach also allows for straightforward setting up of new contracts at runtime, including references to locations not known at compile-time. Furthermore, the migrating approach procrastinates from setting up monitors in remote locations until necessary. In contrast, on a choreographed approach, monitors are set up at all locations, even though some of them may never be triggered.


Formalising the compilation of regular expression contracts into \MDPi\  also gives us opportunities to formally verifying certain properties. For instance, as a generalisation of \eqref{eq:1sem} we can state and prove that,  for arbitrary expression $E$,  different compilation approaches give the same monitoring result.   We can state this as:
$$
\conf{\delta}{\textsl{Sys} \piPar \psi_O(E)})_{\Omega_\textsl{NTg}}  \quad \bisim \quad  (\conf{\delta}{\textsl{Sys} \piPar \psi_C(E)})_{\Omega_\textsl{NTg}} \quad \bisim \quad  (\conf{\delta}{\textsl{Sys} \piPar \psi_M(E)})_{\Omega_\textsl{NTg}}
$$
and prove it by giving witness bisimulations defined by induction on the structure of $E$.  One can prove similar results on the lines of the equivalences given in section~\ref{sec:behav-equiv}.

%
%
%
%
%
%
%

\section{Conclusions}
\label{sec:conclusions}

We have presented a novel process calculus framework in which distributed contract monitoring can be formalised and analysed.  We have shown it to be expressive enough to encode various distributed monitoring strategies. To the best of our knowledge, it is unique in that it traces are first class entities rather than meta-constructs.  We modularly developed various semantics for this calculus, using transition abstraction techniques that enable selective reasoning about 
aspects such as locality of communication and distinctions between monitor and process actions.

 We 
are currently working on an implementation in Erlang \cite{erlang}, guided by the design decisions made for our calculus.   This should give us insight into 
 practical issues, such as that of addressing trust issues when installing monitors and the avoidance of indirect data exposure due to monitoring.  
We are also studying \MDPi\ further, addressing issues such as clock boundaries and real-time operators. As the calculus stands, the monitoring component is non-intrusive, in that it reads system events but does not otherwise interact with it. To handle reparations and enforcements upon contract violation, and to be able to express monitor-oriented programming \cite{chen-rosu-2003-rv} we require potentially intrusive monitoring. We believe that our bisimulation approach can also handle reasoning about monitor intrusiveness.



\bibliographystyle{eptcs}
\bibliography{mastersbiblio}



\end{document}